\documentclass[useAMS,usenatbib]{mnras}

\usepackage{times}
\usepackage{ amssymb }
\usepackage{mathptmx}
\usepackage{multirow}
\usepackage{psfrag}
\usepackage{pspicture}
\usepackage{graphicx}
\usepackage{amsmath}
\usepackage{color}
\usepackage{url}
\usepackage{caption}
\usepackage{subcaption}
\usepackage{grffile}

\title[The optimal gravitational softening length for Cosmological N-body simulations]
  {The optimal gravitational softening length for cosmological N-body simulations}

\author[Zhang et al.]
{Tianchi Zhang,$^{1, 2}$\thanks{Email: tczhang@nao.cas.cn} Shihong Liao,$^1$ Ming Li,$^1$ Liang Gao$^{1, 2, 3}$\\
$^1$Key Laboratory for Computational Astrophysics, National Astronomical Observatories, Chinese Academy of Sciences, Beijing 100012, China\\
$^2$School of Astronomy and Space Science, University of Chinese Academy of Sciences, Beijing 100049, China\\
$^3$Durham University, South Road, Durham, DH1 3LE, UK\\
}
\begin{document} 



\maketitle

\label{firstpage}

\begin{abstract}
Gravitational softening length is one of the key parameters to properly set up a cosmological $N$-body simulation. In this paper, we perform a large suit of high-resolution $N$-body simulations to revise the optimal softening scheme proposed by Power et al. (P03). Our finding is that P03 optimal scheme works well but is over conservative. Using smaller softening lengths than that of P03 can achieve higher spatial resolution and numerically convergent results on both circular velocity and density profiles. However using an over small softening length overpredicts matter density at the inner most region of dark matter haloes. We empirically explore a better optimal softening scheme based on P03 form and find that a small modification works well. This work will be useful for setting up cosmological simulations.

\end{abstract}


\section{Introduction}\label{sec_intro}
Cosmological $N$-body simulations are essential to study the formation of large-scale structures in Universe. In the past decades, with the rapid developments  both in the computational power of supercomputers and the numerical techniques, cosmological $N$-body simulations have played an important role in the studies of hierarchical formation of cold dark matter haloes and the establishment of the standard cosmological model \citep[see][for a review]{Frenk2012}.

For a cosmological simulation code, there are usually quite a few parameters to be chosen in order to properly set up it, gravitational softening length is one of the key parameters. In cosmological $N$-body simulations, in order to avoid close encounters between particles, a small quantity, $\epsilon$, is introduced in the computation of Newtonian gravity, i.e., the Plummer form, $\mathbf{F}_{12}=Gm_1m_2\mathbf{r}_{12}/(\mathbf{r}_{12}^2 + \epsilon^2)^{3/2}$. Here, $G$ is the gravitational constant, $m_1$ and $m_2$ are the masses of two particles, $\mathbf{r}_{12}$ is the position vector from particle 1 to particle 2, and $\epsilon$ is termed as gravitational softening length. In this sense, instead of being a point mass, a particle is treated as a smooth ``ball'' with a volume measured by the softening length. It is not a trivial task to choose an optimal softening length for a numerical simulation in terms of computational cost and force accuracy. In the past decades, many studies have been performed to explore how to choose softening lengths in $N$-body simulations \citep[see e.g.][]{Thomas1992, Merritt1996, Romeo1997, Romeo1998, moore1998, Splinter1998, Athanassoula2000, Knebe2000, Dehnen2001, Fukushige2001, power2003, Zhan2006, Price2007, Iannuzzi2011, Bosch2018}. For a uniform mass resolution cosmological simulation, the softening length is usually set to be a fraction of the mean inter-particle separation. However there is no consensus on the choice of the fraction. In literature, the fraction varies from $1/120$ \citep[e.g.,][]{bol} to $1/10$ \citep[e.g.,][]{Kim2009}. 

Currently, the most widely adopted setting of the optimal softening length in zoom-in $N$-body simulations is suggested by \citet[][]{power2003} (hereafter P03). P03 proposed an optimal choice of softening length based on the argument that the maximum stochastic acceleration caused by close approaching to a single particle, $a_\mathrm{max}=Gm/\epsilon^{2}$, should be less than the minimum mean-field acceleration in a virial halo, $a_\mathrm{min}\approx GM_{200}/r_{200}^{2}$. Here, $M_{200}$ and $r_{200}$ are the virial mass and virial radius of a simulated halo with its mean density inside $r_{200}$ being $200$ times the critical density. This argument sets a lower limit for the softening length which is needed to avoid strong discreteness effects, $\epsilon > \epsilon_\mathrm{acc} \approx r_{200}/\sqrt{N_{200}}$, where $N_{200}$ is the number of particles within the virial radius. P03 further empirically proposed that an optimal softening length is
\begin{equation}\label{eq_p03_opt}
\epsilon_\mathrm{opt,P03} \approx 4\epsilon_\mathrm{acc} = \frac{4r_{200}}{\sqrt{N_{200}}},
\end{equation}
which tends to describe their numerical results well. With this optimal softening, the circular velocity profile of a halo can converge at the radius $r_{conv}$ at a level of better than $10$ percent \citep[][]{Navarro2004}. Here, the convergence radius is estimated by requiring the collisional relaxation time at the convergence radius, $t_{relax}(r_{conv})$, equals to the circular orbital time at the virial radius, $t_{circ}(r_{200})$, i.e., 
\begin{equation}\label{eq_p03_rconv}
\kappa(r_{conv})=\frac{t_{relax}(r_{conv})}{t_{circ}(r_{200})}=\frac{\sqrt{200}}{8}\frac{N(r_{conv})}{\mathrm{ln}N(r_{conv})}\left[\frac{\rho_{crit}}{\bar{\rho}(r_{conv})}\right]^{\frac{1}{2}}=1,
\end{equation}
where $\rho_{crit}$ is the critical density, and $N(r_{conv})$ and $\bar{\rho}(r_{conv})$ are the enclosed number of particles and mean enclosed density within $r_{conv}$, respectively. The proposal of P03 optimal softening scheme has been widely adopted in the settings of many zoom-in simulations such as the Phoenix simulations \citep[][]{Ph}, Auriga simulations \citep[][]{Au}, AGORA simulations \citep[][]{AGORA}, FABLE simulations \citep[]{FABLE}, etc.

Since the proposal of P03 softening scheme, cosmological simulation codes have evolved gradually in recent years, both in force calculation and time integration accuracy. It is interesting to revisit the problem with the most updated codes and with better statistics to see whether P03 optimal softening scheme still holds, and if not, how to improve it.

In this paper, we will revise P03 optimal softening scheme with a set of high-resolution simulations. The paper is structured as follows. In Section \ref{sec_sim}, we describe the details of our simulations and halo samples. In Section \ref{sec_res}, we use a series of high-resolution numerical simulations to test the optimal softening scheme advocated in P03 (Section \ref{result_1}), and propose an improved optimal softening which can achieve higher spatial resolution (Section \ref{result_2}), and discuss the implications of our updated optimal softening length (Section \ref{result}). Our conclusions are presented in Section \ref{sec_con}.

\section{Numerical Simulations} \label{sec_sim}
We use one of the most widely used cosmological simulation codes, \textsc{Gadget-3}, which is an improved version of \textsc{Gadget-2} \citep[]{springel2005}, to perform all our simulations in this study. The cosmological parameters are $\Omega_m=0.3, \Omega_\Lambda=0.7, \sigma_8=0.9, h=0.65$ and $n_s=1.0$. The initial conditions at $z=100$ are generated with the \textsc{N-genic} code with the linear matter power spectrum given in \citet[]{eisenstein1998}. Dark matter haloes in the simulations are identified with the standard friends-of-friends algorithm with a linking length of 0.2 times the mean particle separation \citep[][]{Davis1985}.

In our simulations, the default integration accuracy parameter ErrTolIntAcc is set to 0.025. For the TreePM computation, the force accuracy parameter ErrTolForceAcc is set to 0.0025, and the FFT mesh dimension, PMGRID, is set to be equal to the number of particles in each dimension, $N_p$. Varying these three parameters hardly affect our results present below; see Appendix \ref{app_B} for details.

\textbf{Simulation set I}. To test the optimal softening scheme in P03, Eq. (\ref{eq_p03_opt}), we perform a set of simulations with varying softening lengths fixed in comoving coordinates. Each simulation contains $N_p^3 = 256^3$ dark matter particles in a periodic box with a length $L_\mathrm{box}=10$ $\mathrm{Mpc}/h$ on a side. 

We first run a simulation with a softening length following the usual choice, $1/50$ of the mean inter-particle separation, i.e., $\epsilon_\mathrm{use} = XL_\mathrm{box}/N_p$ with $X=1/50$. The value of $\epsilon_\mathrm{use}$ is $0.78$ $\mathrm{kpc}/h$ here. Then we select the most massive halo and calculate its optimal softening according to Eq. (\ref{eq_p03_opt}) by using its $r_{200} \approx 330$ $\mathrm{kpc}/h$ and $N_{200} \approx 1.6\times 10^6$. The P03 optimal softening length for this halo is $\epsilon_\mathrm{opt, P03}=1.0$ $\mathrm{kpc}/h$, roughly $X=1/40$ of the mean inter-particle separation. Then we re-run the simulation with $\epsilon_\mathrm{use} = \epsilon_\mathrm{opt, P03}$, and a series of softening lengths greater or less than $\epsilon_\mathrm{opt, P03}$, i.e., $X=1/10, 1/25, 1/80, 1/100,$ $1/300$ and $1/500$. As a fiducial one to compare with, we also perform a simulation with 8 times better mass resolution with $N_p^3=512^3$ and 2 times better spatial resolution, and use the same random phases as the lower resolution runs to set up the initial conditions. We will use this set of simulations to test whether P03 optimal softening scheme works the best to resolve the inner structures of the most massive halo in Section \ref{result_1}.

\textbf{Simulation set II}. As we shall see in next section that P03 softening scheme is indeed not most optimal. In order to improve it, we generalize the form of P03 optimal softening scheme by introducing a free parameter, $\alpha$ (see Section \ref{result_2} for details). We explore whether or not we can improve P03 softening scheme in a simple way by varying $\alpha$ in our following numerical simulations.

In order to have better statistics, we perform a set of cosmological simulations with a box size of $L_\mathrm{box} = 33 \mathrm{Mpc}/h$, and focus on the galactic haloes with masses $M_{200}=[5\times 10^{11}, 2\times 10^{12}]M_\odot/h$ with the corresponding virial radii $r_{200} \approx 150\mathrm{kpc}/h$. There are about $160$ haloes in the halo sample in each simulation, and these haloes are stacked to obtain the stacked density and circular velocity profiles. We have performed the simulations with three different resolutions, at each resolution, we run these simulations with five different softening setups, namely $\alpha = 0.5, 1, 2, 3,$ and $4$. Details of simulations are summarized in Table \ref{tab_1}.

\begin{table}
\begin{center}
\begin{tabular}{l c c c c c c }
\hline
Name  &     $\alpha$        &$m_\mathrm{p}$          & $N_{halo}$ & $\bar{N}_{200}$&&$\epsilon$ \\
            &                          &$[\mathrm{M}_\odot/h]$ &                   &                          &&  $[\mathrm{kpc}/h]$ \\
\hline
Fiducial       &4&    $7.31\times 10^{5}$      &164              &1285515               &&  $0.54$                     \\
\hline
\multirow{4}{*}{\centering HighRes}   &1&\multirow{4}{*}{\centering$5.84\times 10^{6}$}&167     									       &160860              &&  $0.38$                     \\
       &2&           				 &167              &159430               &&  $0.76$                     \\
       &3&	      		                  &166              &162509               &&  $1.14$                     \\
       &4&     				 &163              &162216                &&  $1.52$                     \\
\hline
\multirow{5}{*}{\centering MidRes}         &0.5&  \multirow{5}{*}{\centering $4.68\times 10^{7}$}                                        							&165             &19765                &&  $0.54$                     \\
         &1&                                           &165              &20059                &&  $1.03$                     \\
         &2&          				&167             &19708                &&  $2.06$                      \\
         &3&	      		                         &168             &19887                &&  $3.09$                     \\
         &4&     				         &168            &19580                 &&  $4.124$                    \\
\hline
\multirow{5}{*}{\centering LowRes} &    0.5       & \multirow{5}{*}{\centering $3.74\times 10^{8}$}           						  &168              &2463                  &&  $1.52$      \\
      &1&                                       &170              &2440                  &&  $3.03$                     \\
      &2&         				 &167              &2456                  &&  $6.06$                     \\
      &3&	      		                  &164              &2434                  &&  $9.09$                     \\
      &4&     				 &165              &2418                  &&  $12.12$                     \\
\hline
\end{tabular}
\vspace{10pt}
\caption{Details of simulation set II. Here, $m_p$, $N_{halo}$, and $\bar{N}_{200}$ denote the particle mass, number of selected haloes, and the average number of particles inside the selected haloes, respectively.}
\label{tab_1}
\end{center}
\end{table} 

\section{Results}\label{sec_res}

\subsection{Testing P03 optimal softening scheme}\label{result_1}
\begin{figure*} 
\centering\includegraphics[width=500pt]{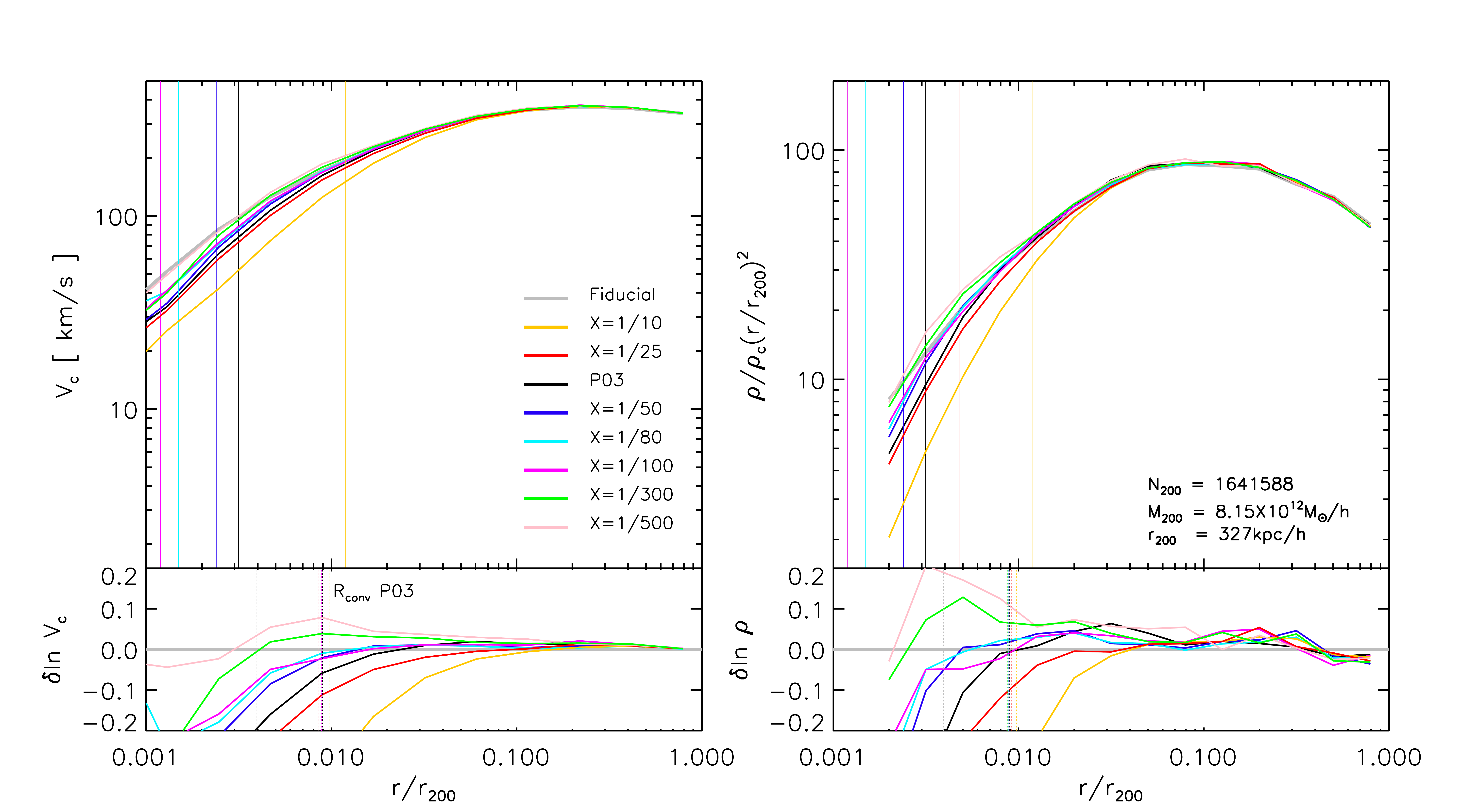} 
\caption{Circular velocity profiles (left panel) and density profiles (right panel) for the largest halo in the simulation set I at $z=0$. For a direct comparison, the radius has been scaled to $r_{200}$ for each halo. In each panel, the upper part presents profiles, and the bottom shows their residuals from the fiducial simulation (e.g., $\delta \ln V_c = \ln V_c - \ln V_{c, Fiducial}$). The solid vertical lines in the upper parts mark the softening lengths for different simulations, whereas the dotted vertical lines in the bottom parts show P03 convergence radii, $r_{conv}$, computed from Eq. (\ref{eq_p03_rconv}).}\label{fig1}
\end{figure*}

To test the optimal softening scheme proposed by P03, in Fig. \ref{fig1}, we plot the circular velocity profile, $V_c(r)=\sqrt{GM(r)/r}$, and density profile, $\rho(r)$, of the most massive halo in each Set I simulation, and compare them with those of the fiducial run. Results for different simulations are distinguished with different colors as labelled in the figure. 

We can clearly see that both circular velocity and density profiles in the simulations with $X=1/80$ and $1/100$ converge to smaller radii when compared with the run using softening length proposed by P03. This suggests that P03 softening scheme may be too conservative \citep[see][for a similar conclusion]{Ludlow2018}. However, we shall note that using an over small softening (e.g. the simulation with $X=1/500$) overpredicts  $\rho(r)$ with respect to the fiducial one as large as $\sim 20\%$ at small radii, this is possibly due to two-body effects introduced by over small softening length, or spurious low-mass structures which form at early times, retain their high central densities and later sink into the halo center to artificially boost the central density \citep[see][for related discussions]{Power2016}. Note that P03 convergence radii (the dotted vertical lines in Fig. \ref{fig1}) are almost independent of the chosen softening lengths. Also we note that, while we only show the results for the most massive halo here,  similar results can be found for other haloes with comparable halo mass in the simulations.

\subsection{An improved proposal of optimal softening and convergence radius}\label{result_2}

As we see in the above subsection that P03 softening scheme is over conservative, For simplicity, in this subsection we explore to improve it based on the original P03 form. To this end, we generalize the form of P03 optimal softening form into \begin{equation} \label{eq_gen_opt}
\epsilon_\mathrm{opt} = \alpha \frac{r_{200}}{\sqrt{N_{200}}},
\end{equation}
where $\alpha$ is a free parameter to be determined here, and $\alpha=4$ corresponds to the original P03 optimal softening scheme. To empirically explore the optimal $\alpha$, we have performed a set of simulations with softening lengths given by $\alpha=0.5, 1, 2, 3$, and $4$, which are described in detail in Section \ref{sec_sim}.

\begin{figure*} 
\centering\includegraphics[width=500pt]{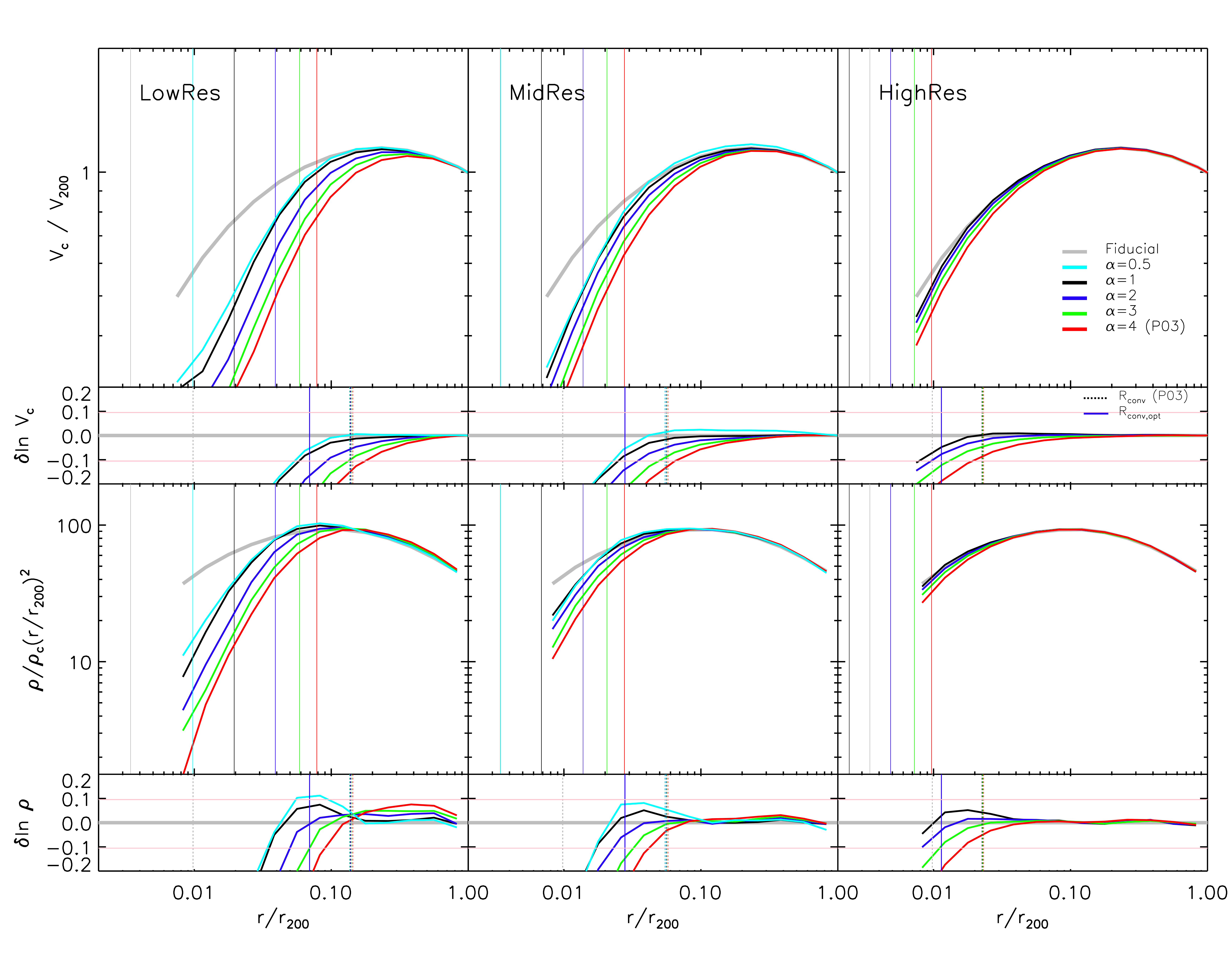}
\caption{Stacked circular velocity and density profiles for galactic haloes at $z=0$ in the simulation set II. Results from the fiducial simulation, the simulations with $\alpha=0.5, 1, 2, 3$, and $4$ are plotted with grey, cyan, black, blue, green and red curves, respectively. From left to right columns, results from LowRes, MidRes and HighRes simulations are shown. The first two rows show circular velocity profiles and their residuals to the fiducial one respectively, while the third and fourth rows show density profiles and their residuals with respect to the fiducial one respectively. The solid vertical lines in the first and third rows mark the softening lengths of simulations. The dotted (solid blue) vertical lines in the second and fourth rows mark the P03 convergence radii (our improved convergence radius estimation, i.e., Eq. (\ref{eq_gen_rconv})), and the horizontal pink in the residual plots lines show $10$ percent for easy reference.}
\label{fig2}
\end{figure*}

To reduce noises, we stack the circular velocity and density profiles of $\sim 160$ galactic haloes at $z=0$ which have masses in the range from $5\times 10^{11}$ $\mathrm{M}_\odot/h$ to $2\times 10^{12}$ $\mathrm{M}_\odot/h$.  In Fig.~\ref{fig2}, we plot the stacked profiles of the simulations with different resolutions and compare them with the fiducial run.

Let us focus on results of MidRes simulations (middle column of Fig. \ref{fig2}) first. Clearly $V_c(r)$ of the simulation using P03 softening scheme (solid red curve) converge to the fiducial one at convergence radius at 10 percent level, in agreement with studies of \citet[][]{Navarro2004} and others. However the simulations with smaller softening length converge to even smaller radii at a similar error level. Results for density profiles are displayed in the lower panels of the same figure. Again, the P03 softening scheme does a pretty good job in matching the density profile at the convergence radius at which density profile of the halo in the lower resolution run only deviates from the fiducial one about few percent. However, similar to the result for the circular velocity profiles, using smaller softening length, the density profile can converge to smaller radii, for $\alpha=2,3$ the spatial resolution for the stacked density profile can be improved by a factor of 1.8 and 1.3, respectively. Note, as we discussed in the last subsection that using an over small softening overpredicts dark matter density at very inner region, one can readily find bumps in the residual plot for the run using $\alpha=0.5, 1$ at $\sim 0.04r_{200}$. Therefore, according to the above convergence tests, these results suggest the simulations with $\alpha=2$ works equivalently well as P03 in terms of numerical convergence but at the same time can achieve about 2 times better spatial resolution. Similar conclusions can also be drawn from LowRes and HighRes simulations.

A remaining question is that if we choose $\alpha=2$ in Eq. (\ref{eq_gen_opt}) as a better optimal softening scheme, then is it possible to give an estimation of its convergence radius? In Fig. \ref{fig2}, we plot P03 convergence radii with vertical dotted lines in the residual panels. Similar to previous studies \citep[e.g.,][]{Navarro2004}, we find that in the MidRes and HighRes cases, the circular velocity profiles with P03 softening (red lines) converge to the fiducial one roughly at a level of $10$ percent at $r = r_{conv, P03}$. But for the LowRes case, the convergence level at $r = r_{conv, P03}$ is slightly worse, i.e., $\sim 15\%$. Note that the haloes in LowRes simulations only have $\sim 2000$ particles, and previous studies \citep[e.g.,][]{Navarro2004} have not tested P03 convergence radius for the haloes with such low number particles. Our LowRes results suggest that in haloes with thousands of particles, the circular velocity profile at the P03 convergence radius converges at a level worse than $10$ percent.

We also plot the half of P03 convergence radius with blue vertical solid lines in the residual panels in Fig. \ref{fig2}. They offer a rough estimation of the convergence radius of the circular velocity with $\alpha = 2$ at a level of $\sim 10\%$. This means that by reducing the softening length into half of P03 optimal softening scheme, the spatial resolution of a simulation can be two times better. In such a way, we efficiently achieve a spatial resolution which otherwise needs a simulation with eight times more particles and several times more computational cost.

We have also looked at a set of simulations targeting cluster haloes, and found similar conclusions for the optimal softening length and convergence radius presented above. Thus, we conclude that an improved proposal for the optimal softening is to set $\alpha=2$ in Eq. (\ref{eq_gen_opt}), and the corresponding convergence radius can be estimated as
\begin{equation} \label{eq_gen_rconv}
r_{conv, opt} = r_{conv, P03}/2,
\end{equation}
where $r_{conv, P03}$ can be computed from Eq. (\ref{eq_p03_rconv}).

\subsection{Discussion}\label{result}

An important application of cosmological simulations is to study the halo mass--concentration relation. In order to estimate concentration parameter ($c$) reliably, simulations need to have enough spatial resolution to well resolve the characteristic radius $r_s$ of a halo of given mass \citep{Neto2007}. Based upon our results presented in the last subsection, we can make a rough estimation of the required mass and spatial resolution in order to reliably estimate the concentration parameter of a halo as a function of halo mass. This will be very useful to set up simulation parameters in practice.

To answer this question, we notice that the enclosed number of particles and mean density in Eq. (\ref{eq_p03_rconv}) can be expressed as
\begin{equation}\label{eq_encl_num}
N(r_{conv,opt}) = \frac{M(r_{conv,opt})}{m_p},
\end{equation}
and
\begin{equation} \label{eq_encl_rho}
\bar{\rho}(r_{conv,opt}) = \frac{3M(r_{conv,opt})}{4\pi r_{conv,opt}^3},
\end{equation}
respectively. Here, $m_p$ is the particle mass, and the enclosed mass within $r_{conv,opt}$ is
\begin{equation}
M(r_{conv,opt}) = M_{200}\frac{f(cx)}{f(c)},
\end{equation}
where $x=r_{conv,opt}/r_{200}$, and the function $f(y)$ has the form of
\begin{equation}
f(y) = \mathrm{ln}(1+y)-\frac{y}{1+y}.
\end{equation}
Considering the relation between $M_{200}$ and $r_{200}$,
\begin{equation} \label{eq_mass_radius}
M_{200} = \frac{800\pi}{3}\rho_{crit}r_{200}^3,
\end{equation}
and putting Eqs (\ref{eq_encl_num}-\ref{eq_encl_rho}) into Eq. (\ref{eq_gen_rconv}), we can find that $r_{conv,opt}$ is a function of $M_{200}$, $c$, and $m_p$. Once $c-M_{200}$ relation is known \citep[e.g.][]{Dutton2014},
$r_{conv,opt}$ is only a function of $M_{200}$ and $m_p$. Therefore, for given $r_{conv,opt}$ and $M_{200}$,  it is easy to derive $m_p$ and then use Eq. (\ref{eq_gen_opt}) and Eq. (\ref{eq_mass_radius}) to compute $\epsilon_\mathrm{opt}$.

\begin{figure} 
\centering\includegraphics[width=250pt]{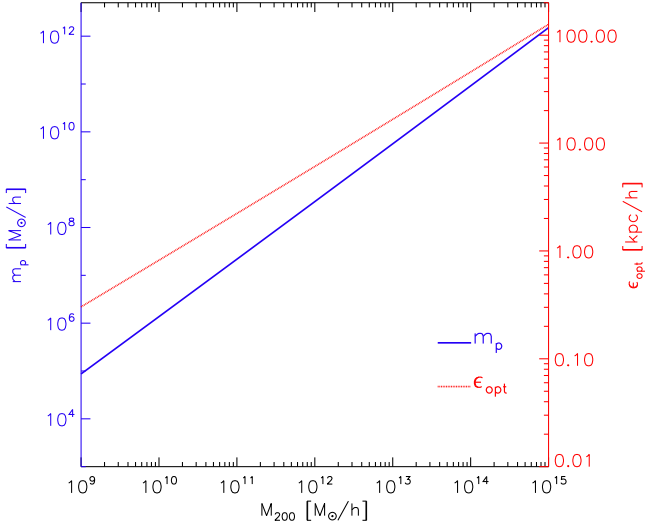} 
\caption{Required mass resolution (left axis) and softening length (right axis) as a function of halo mass assuming $r_{conv,opt}=0.5r_s$. }
\label{fig3}
\end{figure}

In Fig. \ref{fig3}, we plot the required mass resolution $m_p$ (left axis) and optimal softening length as a function of $M_{200}$. The optimal softening length derived here assumes spatial resolution $0.5r_s$ of any given mass halo. From the plot, one can easily identify what mass resolution and softening are needed to reliably estimate the concentration parameter of a halo of given mass when using the optimal softening scheme proposed in this study. For example, if we aim to resolve a Milky Way-sized halo ($M_{200}\sim 10^{12} M_\odot/h$ ), the most economical simulation setup is to use a mass resolution $m_p \approx 5 \times 10^8 M_\odot/h$ and a softening length 5$\mathrm{kpc}/h$, these are indeed very similar to the corresponding parameters adopted in the Millennium simulation \citep[]{ms1}.

\section{Conclusions}\label{sec_con}
We have performed a series of high-resolution cosmological $N$-body simulations to revisit the optimal softening scheme proposed by P03. Our results can be summarized as follows:

(i) We find that P03 optimal softening scheme works well but is over conservative.  Using smaller softening length than the value suggested by P03 can achieve higher spatial resolution and numerically converged results both on circular velocity and density profiles. However using an over small softening causes artificially high density in the inner most of dark matter haloes.

(ii) We empirically generalize the P03 softening scheme by adding a free parameter $\alpha$ (Eq. (\ref{eq_gen_opt})). We use a set of simulations with varying resolutions to show that $\alpha=2$ is an improved choice than the original P03 scheme. We further find that the convergence radius for this updated optimal softening coincides with half of the value in P03.  Therefore, for a given mass resolution, simulations with the improved softening scheme can achieve 2 times better spatial resolution than using P03 one, and thus reduce the computational cost by a large factor for the spatial resolution.

(iii) As the halo mass-concentration relation is an important property to be determined in cosmological simulations, based up our results, we make estimations of the required mass and spatial resolution in order to reliably measure halo concentration parameters as a function of halo mass. 

Our results will be helpful for the set-up of future numerical simulations aiming to study structures of dark matter haloes or galaxies.

\section*{Acknowledgements}
We thank the referee, Chris Power, for an insightful referee report to improve the manuscript. We thank Jie Wang and Qiao Wang for discussions. LG acknowledges support from the national Key Program for Science and Technology Research Development (2017YFB0203300), NSFC grants (Nos 11133003, 11425312) and a Newton Advanced Fellowship, as well as the hospitality of the Institute for Computational Cosmology at Durham University. ML acknowledges support from NSFC grants (No. 11503032), and CPSF-CAS joint Foundation for Excellent Postdoctoral Fellows No. 2015LH0014.

\bibliographystyle{mnras}
\bibliography{paper}         

\appendix
\section{Integration Accuracy, Force Accuracy and PM Grid}\label{app_B}

In this appendix, we study the effects of varying three numerical parameters, integration accuracy, force accuracy and FFT mesh dimension on halo density profiles.

In \textsc{Gadget-3}, the adaptive timestep for a particle is controlled by 
\begin{equation} \label{eq_timestep}
\Delta t=\sqrt{\frac{2\eta\epsilon}{\left|\mathbf{a}\right|}},
\end{equation}
where $\eta$ is the integration accuracy parameter  ErrTolIntAcc, and $\mathbf{a}$ is the particle's acceleration. The default value of $\eta$ for our simulations present in the main text is 0.025.

We adopte the TreePM scheme in \textsc{Gadget-3} to compute gravitational force. For the short-range tree force computation, the relative cell-opening criterion is
\begin{equation}
M l^2>\alpha\left|\mathbf{a}_{old}\right|r^4,
\end{equation}
where $\alpha$ is force accuracy parameter ErrTolForceAcc, $M$ is the mass inside a node, $l$ is cell side-length, $r$ is the distance, and $\mathbf{a}_{old}$ is the total acceleration of the particle. The default value for $\alpha$ is 0.0025. For the long-range PM force, the mesh dimension of the FFT method is given by the parameter PMGRID, and its default value is set to $N_p$.

To examine how these three numerical parameters affect our results, we re-run the $X=1/100$ run the Simulation set I six times more by changing $\eta$, $\alpha$ and PMGRID twice with values 0.5 and 2 times their default respectively. Other parameters and settings of these simulations remain unchanged. The softening length for these testing simulations is chosen to be about the proposed optimal softening lengths of the most massive haloes.

\begin{figure*} 
\centering\includegraphics[width=500pt]{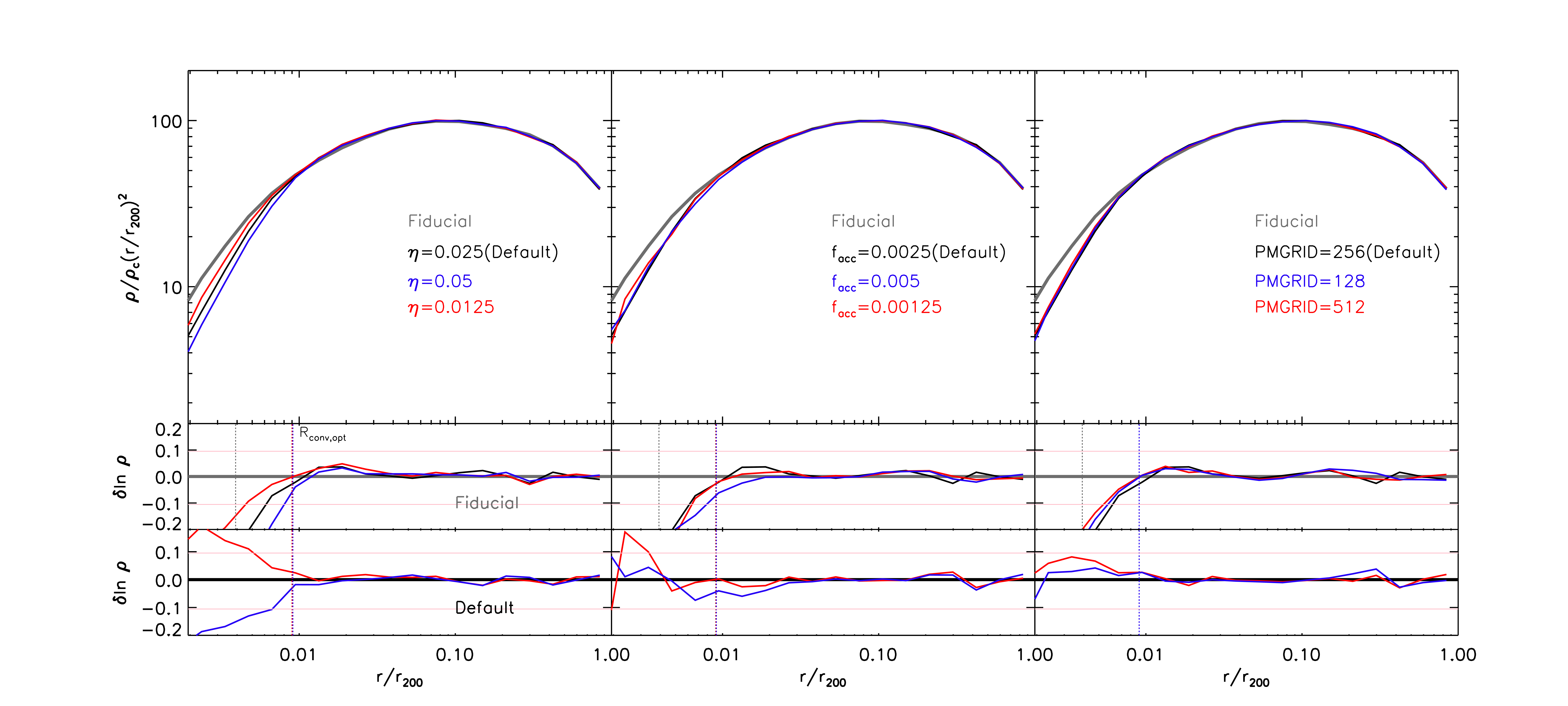} 
\caption{Effects of varying integration accuracy (left), force accuracy (middle) and FFT mesh dimension (right) on halo density profiles. The gray curves represent the fiducial results (i.e. the same as in Fig. \ref{fig1}), while the black, blue and red lines mark the profiles from the simulations with default values, 2 and 0.5 times the default values, respectively. The upper residual panels show the deviations relative to the fiducial curves, while the lower residual panels give the differences with respect to the default results. The vertical dotted lines mark our proposed convergence radii, and the horizontal lines present the 10 per cent convergence region.}\label{figB}
\end{figure*}

To reduce noise, we have stacked the 12 most massive haloes in each simulation, and plot their stacked density profile in Fig. \ref{figB}. As we can see from the bottom residual panels, at radii $r>r_{conv,opt}$, for different $\eta$, $\alpha$ and PMGRID, the changes of halo density profiles are minor (i.e. mostly $\la 5\%$). Especially, when comparing the curves from the simulations with the default values to those with half of the default values, the differences are $\la 2\%$. Therefore, we expect that our results present in main text are not sensitive to the selection of these three parameters.

\bsp	
\label{lastpage}
\end{document}